\documentclass[10pt]{article}
\PassOptionsToPackage{numbers,sort&compress}{natbib}
\usepackage[T1]{fontenc}
\usepackage[utf8]{inputenc}
\makeatletter
\newif\if@neuripsfinal\@neuripsfinalfalse
\newif\if@preprint\@preprinttrue
\newif\if@anonymous\@anonymousfalse
\newif\if@natbib\@natbibtrue
\newcommand{\@neuripsordinal}{40th}
\newcommand{\@neuripsyear}{2026}
\newcommand{\@neuripslocation}{Sydney}

\usepackage{environ}
\newcommand{\acksection}{\section*{Acknowledgments and Disclosure of Funding}}
\NewEnviron{ack}{%
  \acksection
  \BODY
}

\if@natbib
  \RequirePackage{natbib}
\fi

\usepackage[verbose=true,letterpaper]{geometry}
\AtBeginDocument{
  \newgeometry{
    textheight=9in,
    textwidth=5.5in,
    top=1in,
    headheight=12pt,
    headsep=25pt,
    footskip=30pt
  }
  \@ifpackageloaded{fullpage}
    {\PackageWarning{neurips_2026}{fullpage package not allowed! Overwriting formatting.}}
    {}
}

\renewcommand{\normalsize}{%
  \@setfontsize\normalsize\@xpt\@xipt
  \abovedisplayskip      7\p@ \@plus 2\p@ \@minus 5\p@
  \abovedisplayshortskip \z@ \@plus 3\p@
  \belowdisplayskip      \abovedisplayskip
  \belowdisplayshortskip 4\p@ \@plus 3\p@ \@minus 3\p@
}
\normalsize
\renewcommand{\small}{%
  \@setfontsize\small\@ixpt\@xpt
  \abovedisplayskip      6\p@ \@plus 1.5\p@ \@minus 4\p@
  \abovedisplayshortskip \z@  \@plus 2\p@
  \belowdisplayskip      \abovedisplayskip
  \belowdisplayshortskip 3\p@ \@plus 2\p@   \@minus 2\p@
}
\renewcommand{\footnotesize}{\@setfontsize\footnotesize\@ixpt\@xpt}
\renewcommand{\scriptsize}{\@setfontsize\scriptsize\@viipt\@viiipt}
\renewcommand{\tiny}{\@setfontsize\tiny\@vipt\@viipt}
\renewcommand{\large}{\@setfontsize\large\@xiipt{14}}
\renewcommand{\Large}{\@setfontsize\Large\@xivpt{16}}
\renewcommand{\LARGE}{\@setfontsize\LARGE\@xviipt{20}}
\renewcommand{\huge}{\@setfontsize\huge\@xxpt{23}}
\renewcommand{\Huge}{\@setfontsize\Huge\@xxvpt{28}}

\renewcommand{\tiny}{\fontsize{6pt}{7pt}\selectfont}

\renewcommand{\scriptsize}{\fontsize{7pt}{8pt}\selectfont}

\renewcommand{\footnotesize}{\fontsize{8pt}{9.5pt}\selectfont}

\providecommand{\section}{}
\renewcommand{\section}{%
  \@startsection{section}{1}{\z@}%
                {-2.0ex \@plus -0.5ex \@minus -0.2ex}%
                { 1.5ex \@plus  0.3ex \@minus  0.2ex}%
                {\large\bf\raggedright}%
}
\providecommand{\subsection}{}
\renewcommand{\subsection}{%
  \@startsection{subsection}{2}{\z@}%
                {-1.8ex \@plus -0.5ex \@minus -0.2ex}%
                { 0.8ex \@plus  0.2ex}%
                {\normalsize\bf\raggedright}%
}
\providecommand{\subsubsection}{}
\renewcommand{\subsubsection}{%
  \@startsection{subsubsection}{3}{\z@}%
                {-1.5ex \@plus -0.5ex \@minus -0.2ex}%
                { 0.5ex \@plus  0.2ex}%
                {\normalsize\bf\raggedright}%
}
\providecommand{\paragraph}{}
\renewcommand{\paragraph}{%
  \@startsection{paragraph}{4}{\z@}%
                {1.5ex \@plus 0.5ex \@minus 0.2ex}%
                {-1em}%
                {\normalsize\bf}%
}
\providecommand{\subparagraph}{}
\renewcommand{\subparagraph}{%
  \@startsection{subparagraph}{5}{\z@}%
                {1.5ex \@plus 0.5ex \@minus 0.2ex}%
                {-1em}%
                {\normalsize\bf}%
}

\newlength{\@neuripsabovecaptionskip}
\newlength{\@neuripsbelowcaptionskip}

\renewenvironment{table}
  {\setlength{\abovecaptionskip}{\@neuripsbelowcaptionskip}%
   \setlength{\belowcaptionskip}{\@neuripsabovecaptionskip}%
   \@float{table}}
  {\end@float}

\renewcommand{\footnoterule}{\kern-3\p@ \hrule width 12pc \kern 2.6\p@}
\def\@listi  {\leftmargin\leftmargini}
\def\@listii {\leftmargin\leftmarginii
              \labelwidth\leftmarginii
              \advance\labelwidth-\labelsep
              \topsep  2\p@ \@plus 1\p@    \@minus 0.5\p@
              \parsep  1\p@ \@plus 0.5\p@ \@minus 0.5\p@
              \itemsep \parsep}
\def\@listiii{\leftmargin\leftmarginiii
              \labelwidth\leftmarginiii
              \advance\labelwidth-\labelsep
              \topsep    1\p@ \@plus 0.5\p@ \@minus 0.5\p@
              \parsep    \z@
              \partopsep 0.5\p@ \@plus 0\p@ \@minus 0.5\p@
              \itemsep \topsep}
\def\@listiv {\leftmargin\leftmarginiv
              \labelwidth\leftmarginiv
              \advance\labelwidth-\labelsep}
\def\@listv  {\leftmargin\leftmarginv
              \labelwidth\leftmarginv
              \advance\labelwidth-\labelsep}
\def\@listvi {\leftmargin\leftmarginvi
              \labelwidth\leftmarginvi
              \advance\labelwidth-\labelsep}

\providecommand{\maketitle}{}
\renewcommand{\maketitle}{%
  \par
  \begingroup
    \renewcommand{\thefootnote}{\fnsymbol{footnote}}
    \renewcommand{\@makefnmark}{\hbox to \z@{$^{\@thefnmark}$\hss}}
    \long\def\@makefntext##1{%
      \parindent 1em\noindent
      \hbox to 1.8em{\hss $\m@th ^{\@thefnmark}$}##1
    }
    \thispagestyle{empty}
    \@maketitle
    \@thanks
    \@notice
  \endgroup
  \let\maketitle\relax
  \let\thanks\relax
}

\newcommand{\@toptitlebar}{
  \hrule height 4\p@
  \vskip 0.25in
  \vskip -\parskip%
}
\newcommand{\@bottomtitlebar}{
  \vskip 0.29in
  \vskip -\parskip
  \hrule height 1\p@
  \vskip 0.09in%
}

\providecommand{\@maketitle}{}
\renewcommand{\@maketitle}{%
  \vbox{%
    \hsize\textwidth
    \linewidth\hsize
    \vskip 0.1in
    \@toptitlebar
    \centering
    {\LARGE\bf \@title\par}
    \@bottomtitlebar
    \if@anonymous
      \begin{tabular}[t]{c}\bf\rule{\z@}{24\p@}
        Anonymous Author(s) \\
        Affiliation \\
        Address \\
        \texttt{email} \\
      \end{tabular}%
    \else
      \def\And{%
        \end{tabular}\hfil\linebreak[0]\hfil%
        \begin{tabular}[t]{c}\bf\rule{\z@}{24\p@}\ignorespaces%
      }
      \def\AND{%
        \end{tabular}\hfil\linebreak[4]\hfil%
        \begin{tabular}[t]{c}\bf\rule{\z@}{24\p@}\ignorespaces%
      }
      \begin{tabular}[t]{c}\bf\rule{\z@}{24\p@}\@author\end{tabular}%
    \fi
    \vskip 0.3in \@minus 0.1in
  }
}

\newcommand{\ftype@noticebox}{8}
\newcommand{\@notice}{%
  \enlargethispage{2\baselineskip}%
  \@float{noticebox}[b]%
    \footnotesize\@noticestring%
  \end@float%
}

\renewenvironment{abstract}%
{%
  \vskip 0.075in%
  \centerline%
  {\large\bf Abstract}%
  \vspace{0.5ex}%
  \begin{quote}%
}
{
  \par%
  \end{quote}%
  \vskip 1ex%
}

\newcommand{\answerTODO}[1][]{\textcolor{red}{\bf [TODO]}}
\newcommand{\justificationTODO}[1][]{\textcolor{red}{\bf [TODO]}}

\if@preprint
  \newcommand{\@noticestring}{%
    Preprint.%
  }
\else
  \if@neuripsfinal
    \newcommand{\@noticestring}{
      \@trackname
    }
  \else
    \newcommand{\@noticestring}{%
      Submitted to \@neuripsordinal\/ Conference on Neural Information Processing Systems (NeurIPS \@neuripsyear). Do not distribute.%
    }

    \NewEnviron{hide}{}

    \RequirePackage{lineno}
    \linenumbers

    \AtBeginDocument{%
      \@ifpackageloaded{amsmath}{%
        \newcommand*\patchAmsMathEnvironmentForLineno[1]{%
          \expandafter\let\csname old#1\expandafter\endcsname\csname #1\endcsname
          \expandafter\let\csname oldend#1\expandafter\endcsname\csname end#1\endcsname
          \renewenvironment{#1}%
                          {\linenomath\csname old#1\endcsname}%
                          {\csname oldend#1\endcsname\endlinenomath}%
        }%
        \newcommand*\patchBothAmsMathEnvironmentsForLineno[1]{%
          \patchAmsMathEnvironmentForLineno{#1}%
          \patchAmsMathEnvironmentForLineno{#1*}%
        }%
        \patchBothAmsMathEnvironmentsForLineno{equation}%
        \patchBothAmsMathEnvironmentsForLineno{align}%
        \patchBothAmsMathEnvironmentsForLineno{flalign}%
        \patchBothAmsMathEnvironmentsForLineno{alignat}%
        \patchBothAmsMathEnvironmentsForLineno{gather}%
        \patchBothAmsMathEnvironmentsForLineno{multline}%
      }
      {}
    }
  \fi
\fi

\renewcommand{\@noticestring}{Preprint. September 2026.}
\makeatother
\usepackage{amsmath,amssymb,booktabs,tabularx,array,placeins}
\usepackage{graphicx,tikz,pgfplots,listings,xurl}
\usetikzlibrary{arrows.meta,positioning}
\pgfplotsset{compat=1.18}
\usepackage[hidelinks]{hyperref}
\setcitestyle{numbers,square,comma}
\hypersetup{
  pdftitle={Let Execution Do the Teaching},
  pdfauthor={Abhishek Mishra, Sandeep Srinivasa, Arko Chattopadhyay, Enrique Martinho Ferrao, Guda Pranav Reddy},
  pdfsubject={Execution-guided post-training of coding models}
}
\title{Let Execution Do the Teaching\\[4pt]
\large\normalfont\itshape How far can we push a coding model when execution, not humans, does the grading?}
\author{%
  Abhishek Mishra\qquad Sandeep Srinivasa\\[4pt]
  \textbf{Arko Chattopadhyay\qquad Enrique Martinho Ferrao\qquad Guda Pranav Reddy}%
}
\date{}
\begin{document}
\maketitle
\begin{abstract}
Improving a coding model is a sequence of decisions: find the limiting behavior, create useful practice, and determine which attempts deserve reinforcement. We investigate a practical way to make those decisions increasingly repeatable by grounding feedback in execution. Starting from GLM-4.7-Flash, the work combines whole-file supervised fine-tuning, targeted C++ repair curricula, group-relative reinforcement learning, and verifiers that turn program behavior into structured rewards. Compiler errors and failing tests provide specific contexts for repair practice, while executable checks score the resulting programs. We evaluate the trained models on 26 C++ coding tasks, measuring both initial success and completion after test feedback. The strongest reported post-training configuration achieves 45.2\% success in the first test round and 61.5\% within two rounds. These results come from an assisted development suite with evolving prompt contracts; the strongest configuration uses four selected evaluation trials and includes six tasks also used in training. The study describes an execution-guided workflow for improving coding models, including how repair tasks are constructed, how program behavior determines rewards, and how the resulting models are evaluated.
\end{abstract}

\section{Introduction}

A generated C++ implementation can fail during compilation or when tested against the task's requirements. For example, a phone-number cleaner may accept a number whose exchange begins with 1, violating the task specification. Another implementation may fail to compile because a data member and a method share the same name. The first failure requires a change to input validation; the second requires a correction to the C++ interface.

We examine whether execution-graded practice improves a coding model's initial task completion and its ability to repair a failed attempt using test feedback. Compiler diagnostics and test failures identify the errors addressed by the repair tasks. Executable checks then score candidate programs against the task requirements.

We study GLM-4.7-Flash~\cite{glmflash} in a whole-file editing workflow. The model receives a task and source files, then returns complete replacement file contents. A successful response must follow the editing interface, compile, and pass the behavioral checks. Evaluation distinguishes response-format errors, compilation failures, and incorrect program behavior.

Supervised fine-tuning~\cite{flan} trains the model on full solutions and prepared repair examples. Reinforcement learning samples candidate solutions and uses their execution-derived rewards to guide updates. We use low-rank adapters~\cite{lora} for these updates. The verifiers score individual behavioral requirements, and recorded execution outcomes make those scores inspectable. Evaluations of saved checkpoints measure initial task completion and completion after test feedback.

\clearpage
\begin{figure}[!t]
\centering
\definecolor{overviewink}{HTML}{243341}
\definecolor{overviewblue}{HTML}{3568A6}
\definecolor{overviewteal}{HTML}{257D73}
\definecolor{overviewamber}{HTML}{A96D2A}
\definecolor{overviewmuted}{HTML}{596875}
\begin{tikzpicture}[x=1cm,y=1cm,
  font=\sffamily\footnotesize,text=overviewink,
  conn/.style={-{Stealth[length=4pt,width=3.2pt]},line width=0.65pt,draw=overviewmuted},
  card/.style={rounded corners=2.5pt,fill=white,line width=0.55pt,
    text width=3.40cm,minimum height=1.20cm,align=center,inner sep=4pt},
  traincard/.style={card,text width=2.65cm,minimum height=1.20cm,draw=overviewteal!45},
  heading/.style={anchor=west,font=\sffamily\small\bfseries},
  output/.style={font=\sffamily\scriptsize,text=overviewmuted,align=center},
  badge/.style={circle,minimum size=0.47cm,inner sep=0pt,text=white,font=\sffamily\footnotesize\bfseries}]

% Stage panels keep connectors in dedicated gutters.
\draw[rounded corners=5pt,draw=overviewblue!30,fill=overviewblue!5,line width=0.55pt]
  (0,0) rectangle (13.70,-2.50);
\node[badge,fill=overviewblue] at (0.55,-0.43) {1};
\node[heading,text=overviewblue] at (0.98,-0.43) {Task construction and verifier design};
\node[card,draw=overviewblue!45] (failure) at (2.30,-1.38)
  {\textbf{Observed failures}\\Compiler diagnostics\\Failing behavioral tests};
\node[card,draw=overviewblue!45] (practice) at (6.85,-1.38)
  {\textbf{Practice contexts}\\Full-solution examples\\Prepared repairs};
\node[card,draw=overviewblue!45] (validation) at (11.40,-1.38)
  {\textbf{Validate exercises}\\Reference solutions\\Fault controls and checks};
\draw[conn,draw=overviewblue] (failure) -- (practice);
\draw[conn,draw=overviewblue] (practice) -- (validation);
\node[output] at (6.85,-2.17)
  {Training package: examples, repair prompts, and executable checks};

\draw[conn] (6.85,-2.50) -- (6.85,-3.05);
\draw[rounded corners=5pt,draw=overviewteal!30,fill=overviewteal!5,line width=0.55pt]
  (0,-3.05) rectangle (13.70,-6.38);
\node[badge,fill=overviewteal] at (0.55,-3.48) {2};
\node[heading,text=overviewteal] at (0.98,-3.48) {Model training};
\node[traincard] (sft) at (1.80,-4.48)
  {\textbf{SFT warm start}\\Task-answer pairs\\Whole-file edits};
\node[traincard] (generate) at (5.17,-4.48)
  {\textbf{Generate edits}\\Model + adapter\\Whole-file edits};
\node[traincard] (execute) at (8.54,-4.48)
  {\textbf{Execute and score}\\Parse, compile, test\\Behavioral rewards};
\node[traincard] (update) at (11.91,-4.48)
  {\textbf{GRPO update}\\Relative rewards\\LoRA parameters};
\draw[conn,draw=overviewteal] (sft) -- (generate);
\draw[conn,draw=overviewteal] (generate) -- (execute);
\draw[conn,draw=overviewteal] (execute) -- (update);
\draw[conn,draw=overviewteal,rounded corners=4pt]
  (update.south) -- (11.91,-5.65) -- (5.17,-5.65) -- (generate.south);
\node[fill=overviewteal!5,inner sep=3pt,font=\sffamily\scriptsize,text=overviewteal]
  at (8.54,-5.65) {Sample again with the updated adapter};
\node[output] at (6.85,-6.08)
  {Saved adapter checkpoints and recorded execution outcomes};

\draw[conn] (6.85,-6.38) -- (6.85,-6.93);
\draw[rounded corners=5pt,draw=overviewamber!30,fill=overviewamber!5,line width=0.55pt]
  (0,-6.93) rectangle (13.70,-9.43);
\node[badge,fill=overviewamber] at (0.55,-7.36) {3};
\node[heading,text=overviewamber] at (0.98,-7.36) {Checkpoint evaluation};
\node[card,draw=overviewamber!45] (checkpoint) at (2.30,-8.31)
  {\textbf{Saved checkpoint}\\Recorded adapter\\iteration};
\node[card,draw=overviewamber!45] (evaluation) at (6.85,-8.31)
  {\textbf{26 C++ tasks}\\First tests, test feedback,\\and second tests};
\node[card,draw=overviewamber!45] (result) at (11.40,-8.31)
  {\textbf{Record results}\\First-round success\\Two-round completion};
\draw[conn,draw=overviewamber] (checkpoint) -- (evaluation);
\draw[conn,draw=overviewamber] (evaluation) -- (result);
\node[output] at (6.85,-9.10)
  {Development-suite results: trial selection and training-task overlap apply};

% Show the human decision explicitly without a perimeter-spanning return wire.
\draw[conn,dashed] (6.85,-9.43) -- (6.85,-9.92);
\node[draw=overviewmuted!40,dashed,rounded corners=3pt,fill=white,
  text width=11.25cm,minimum height=0.65cm,inner sep=3pt,align=center,
  font=\sffamily\footnotesize,text=overviewmuted]
  at (6.85,-10.25)
  {Researcher reviews the failures and selects the next intervention};
\end{tikzpicture}
\caption{Execution-guided post-training overview. Blue groups task construction and validation, green shows model training, and ochre shows checkpoint evaluation. Within the training loop, execution-derived rewards guide GRPO adapter updates. The dashed handoff marks the researcher's selection of the next intervention. This schematic summarizes components across experiments with different training lineages and reward configurations; the evaluation qualifications in Section~\ref{sec:contracts} apply.}
\label{fig:workflow}
\end{figure}
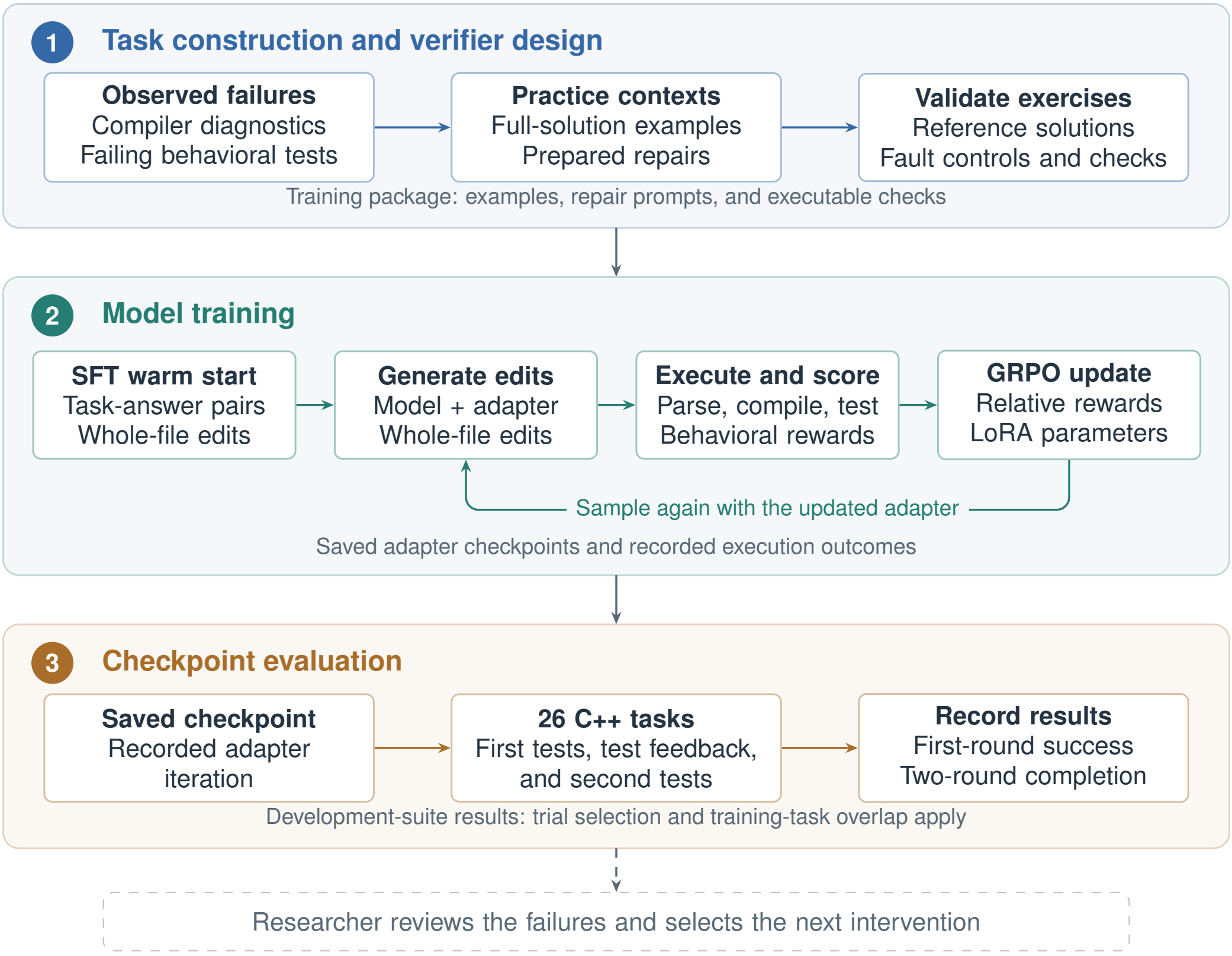

\section{Related Work}

\textbf{Execution-derived supervision.} LETI fine-tunes code generators with a language-modeling objective on sequences conditioned on execution feedback, including diagnostic text and reward tokens indicating whether a solution is correct. \cite{leti} Our reinforcement-learning experiments use diagnostic context in prepared repair prompts and scalar rewards computed from executable checks. The diagnostic context specifies the failure to repair; the reward measures the behavior of the resulting program.

\textbf{Executable task construction.} SWE-smith creates repository-level training tasks by introducing changes that cause previously passing tests to fail, then generates issue descriptions from the code changes and test evidence. \cite{swesmith} Our curricula use C++ exercises with specified interfaces and editable files, including full-solution tasks and prepared repairs targeting named failure mechanisms. The bank-account curriculum checks reference and faulty implementations before packaging its exercises.  The comparison concerns how executable repair tasks are constructed and validated; this study does not evaluate repository-scale task generation.

\textbf{Policy-dependent curricula.} FrogNano's TaskPilot uses current-policy rollouts to admit, revise, and re-evaluate generated repository tasks. \cite{frognano} Our generalized curriculum also uses observed policy success for task selection: an admitted task must have one to three first-round successes across four recorded trials from a specified checkpoint.  This rule selects among prepared tasks. TaskPilot additionally uses rollout feedback to revise the tasks themselves during iterative synthesis.

\textbf{Editing interfaces.} SWE-agent examines how tools for repository navigation, editing, and execution affect performance while holding the underlying model fixed. \cite{sweagent} Our study trains a model to return complete replacement contents for allowed source files, which are then parsed, compiled, and tested.  The reported checkpoint comparisons include changes to the evaluation prompts and supplied API/build information. They therefore do not isolate the effect of model updates from changes in the evaluation setup.

\section{Methodology}

\subsection{Failure mechanisms and repair contexts}

We construct targeted repair tasks for specific failure mechanisms. For example, a C++ phone-number validation implementation failed to compile because its private data member \nolinkurl{number} collided with its public \nolinkurl{number()} method. The repair prompt includes the faulty header and a diagnosis of the name collision. The model must produce an implementation that compiles against the required public interface.

Another repair task targets missing checks on North American area and exchange prefixes. A candidate can compile while accepting numbers with invalid area or exchange prefixes. The repair prompt pairs defective source with the requirement that prefixes beginning with 0 or 1 raise an error. The two repairs address different failures: the C++ declaration collision and the missing area- and exchange-prefix checks.

\begin{table}[htbp]\centering
\caption{Phone Number inputs, requirements, and targeted repair mechanisms.}
\label{tab:t1}
\small\setlength{\tabcolsep}{4pt}\renewcommand{\arraystretch}{1.14}
\begin{tabularx}{\linewidth}{@{}>{\hsize=1.03\hsize\linewidth=\hsize\raggedright\arraybackslash}X>{\hsize=1.12\hsize\linewidth=\hsize\raggedright\arraybackslash}X>{\hsize=0.85\hsize\linewidth=\hsize\raggedright\arraybackslash}X@{}}
\toprule

\textbf{Concrete input or observation} & \textbf{Required behavior} & \textbf{Lesson} \\ \midrule

+1 (223) 456-7890 & Return 2234567890; render (223) 456-7890. & Preserve valid normalization. \\

2231567890 & Throw \nolinkurl{std::domain_error}. & Validate the exchange prefix. \\

Member named \nolinkurl{number} and method \nolinkurl{number()} & Produce an interface that compiles. & Repair the declaration collision. \\

\bottomrule
\end{tabularx}
\end{table}

\subsection{Curriculum construction}

The task builder emits eight prompts for this example: one full-solution task and seven repair contexts. The repairs cover missing definitions, invalid prefixes, character handling, length, country code, display formatting, and the observed name collision. All eight prompts specify the same public interface and editable files. The model generates complete replacement contents for those files, and the runner checks the resulting program.

\subsection{Use execution to check both the exercise and the answer}

The task builder validates exercises with reference implementations and faulty controls. In the bank-account example, it introduces a compilation or linking fault and a state-management fault that retains the old balance after reopening. A separate boundary fault accepts a zero amount. Before packaging the exercises, the builder requires the reference implementation to pass and each faulty candidate to fail at its expected stage.

For the phone-number validation exercise, the verifier compiles the candidate against the required C++17 interface and runs twelve Boolean checks. These checks cover normalization, invalid characters, length and country-code partitions, prefix restrictions, and public output behavior. A \emph{kernel} is one scored Boolean check and may contain several test inputs. The implementation organizes the twelve checks into four groups of three; Table~\ref{tab:phonegroups} summarizes their coverage.

\begin{table}[htbp]
\centering
\caption{Coverage of the four verification groups. Descriptive labels summarize the checks; identifiers match the released implementation. Each group contains three Boolean kernels. }
\label{tab:phonegroups}
\small\setlength{\tabcolsep}{4pt}\renewcommand{\arraystretch}{1.14}
\begin{tabularx}{\linewidth}{@{}>{\hsize=0.76\hsize\linewidth=\hsize\raggedright\arraybackslash}X>{\hsize=1.24\hsize\linewidth=\hsize\raggedright\arraybackslash}X@{}}
\toprule
\textbf{Group} & \textbf{Checks} \\ \midrule
Basic valid-input behavior (\nolinkurl{PH-E01}) & Plain ten-digit input, area-code and number accessors, and a leading country code of 1. \\
Formatting and rejection rules (\nolinkurl{PH-E02}) & Permitted formatting; invalid length, country code, and characters; invalid area and exchange prefixes. \\
Formatting and boundary cases (\nolinkurl{PH-E03}) & Whitespace, the valid number 9999999999, exact string formatting, and empty, overlong, or malformed inputs. \\
Combined interface and input checks (\nolinkurl{PH-E04}) & Consistent public outputs, multiple valid formats, and combined rejection cases. \\
\bottomrule
\end{tabularx}
\end{table}

\FloatBarrier

\paragraph{From twelve outcomes to one reward.}

For checks $b_1,\ldots,b_{12}$, each equal to 0 or 1:

\begin{equation}
R=\frac{1}{12}\sum_{k=1}^{12}(2b_k-1).
\end{equation}

All twelve passing gives +1. Ten passing gives +2/3. All twelve failing gives -1. These are arithmetic examples of the reward rule, not sampled experimental outcomes.

Programs receive different rewards according to how many checks they pass. Recorded check outcomes show which requirements a program passes or fails.

The verifier writes the twelve outcomes as a 12-bit record in the test log. The reward parser accepts that record only if the harness's completion marker directly follows it. The marker contains a fresh random token for the run. If the harness reports a passing result without the expected record, the reward path flags the sample as an infrastructure error rather than accepting it as a pass.

\subsection{Task admission}

If every attempt receives the same reward, that group offers no reward preference among its samples. The multi-task C++ curriculum admits a task only if it has one to three first-round successes across four recorded trials from the specified source checkpoint. The admission procedure verifies the receipt file hashes and their checkpoint, run, and task-suite identifiers. It recomputes trial and success counts and compares them with the admission record.

\subsection{GRPO training and checkpoint evaluation}

GRPO~\cite{deepseekmath} samples several responses to the same prompt and uses their relative rewards to guide the update. In the phone-number validation run, eight prompts each produced 32 samples, yielding 256 responses per rollout batch. We trained a LoRA adapter and saved it for checkpoint evaluation. \cite{lora} After training, checkpoint evaluation measured first-round solving and repair after test feedback.

\section{Engineering}

\subsection{Response parsing and file validation}

The parser extracts filename-labeled code blocks from the final answer after any reasoning text. It checks each target against the allowed file set. The parser accepts Markdown-wrapped labels and path labels with an allowed basename but records them as formatting errors. In strict mode, duplicate file blocks and targets outside the allowed set cause rejection. With boundary recovery enabled, the parser skips forbidden targets and keeps the last duplicate block while marking the response as format-invalid. The runner compiles the candidate from the parsed file edits.

\subsection{Infrastructure error handling}

The reward path distinguishes candidate failures from execution-environment errors. It retains the raw execution reward and diagnostic flags separately from the adjusted training score. When using this adjustment, the reward path groups samples by rollout and problem. Each infrastructure-invalid sample receives the mean training score of the valid samples in its group. If the group contains no valid samples, every sample receives a zero training score. Assigning the valid-group mean removes the invalid sample's deviation from the mean after group centering. The original reward and error flags remain in the record.

\subsection{Adapter loading and synchronization}

After loading previously trained adapter weights, the integration refreshes the optimizer's FP32 master weights to match the loaded adapter. When training offload and LoRA are enabled, it restores saved CPU adapter snapshots to their original devices before the weight updater synchronizes the rollout service.

\subsection{Shared C++ verification}

The shared verifier checks source structure and API requirements, separate compilation and linking, official-test completion, response integrity, and candidate-output completeness. The code calls the five-group profile \nolinkurl{live}. The \nolinkurl{full} profile adds warning and sanitizer diagnostics and requires all selected checks to pass for a profile-level pass. Response checks examine repeated text and incomplete code blocks. Candidate test runs must complete without crashes or timeouts and satisfy the required tests.

Before awarding a positive verification reward, the implementation validates the check records and the reference and candidate execution results. It requires successful reference execution, verified candidate completion, zero required return codes, and no execution-failure flags. The runner also checks that candidate source files remain unchanged across verification. A validated pass earns +1, or +0.975 when a recoverable formatting error remains. Incomplete candidates receive scores between -1 and 0. Their partial credit comes from weighted passed-check results and the available semantic test fraction. The implementation flags invalid verification records as infrastructure errors.

For the dice-scoring exercise, Yacht, the checker enumerates all 7,776 ordered rolls of five six-sided dice across twelve scoring categories. For the list-comparison exercise, Sublist, it tests equality and contiguous sublist or superlist relationships. Its exhaustive check covers 115,600 ordered pairs of nonempty lists of lengths one through four over the values $\{-1,0,1,2\}$. The checker handles empty-list cases separately. For the perfect-number classification exercise, the reference implementation sums proper divisors using a 64-bit accumulator and divisor pairs. A separate checker computes the sum through prime factorization. Its tests include inputs from 1 through 4,096 and large values near the 32-bit integer limit.

For tasks with required task-specific checks, the combined reward stays positive only when both shared and task-specific verification pass. Failure in either produces a non-positive score. The implementation flags invalid task-specific results as infrastructure errors.

This section describes the released verifier implementation. The checkpoint experiments in Section~\ref{sec:experiments} use the reward and evaluation configurations recorded for their own runs. The source tracks later verifier changes separately; those changes are not retroactively part of these results.

\section{Experiments}

\label{sec:experiments}

\subsection{Evaluation protocol}

Fixed26 is the project's fixed set of 26 C++ coding tasks evaluated through Aider~\cite{aider}. Four trials per reported configuration produce 104 task-trial trajectories. The first-round metric asks whether a task passes its first test round. The cumulative two-round metric asks whether it is solved by the end of the second test round, after feedback. A test round may contain multiple model requests.

Conditional recovery is the fraction of first-round failures that become successes in round two. Keeping this denominator visible helps distinguish a model that starts more reliably from one that repairs more effectively.

\begin{table}[htbp]\centering
\caption{Evaluation metrics and their denominators.}
\label{tab:t2}
\small\setlength{\tabcolsep}{4pt}\renewcommand{\arraystretch}{1.14}
\begin{tabularx}{\linewidth}{@{}>{\hsize=0.8\hsize\linewidth=\hsize\raggedright\arraybackslash}X>{\hsize=1.1\hsize\linewidth=\hsize\raggedright\arraybackslash}X>{\hsize=1.1\hsize\linewidth=\hsize\raggedright\arraybackslash}X@{}}
\toprule

\textbf{Metric} & \textbf{Calculation across four trials} & \textbf{Interpretation} \\ \midrule

First-round mean & First-round successes / 4 & Tasks solved immediately, out of 26. \\

Two-round mean & Tasks solved by round two / 4 & Cumulative completion, out of 26. \\

Conditional recovery & New round-two successes / first-round failures & Use of the available feedback opportunity. \\

\bottomrule
\end{tabularx}
\end{table}

\subsection{Evaluated configurations and prompt contracts}

\label{sec:contracts}

The results include base GLM-4.7-Flash, two supervised configurations, two execution-midband RL configurations, Phone Number GRPO, and Generalized C++ GRPO. SFT v5 uses 1,117 Aider-format examples over three epochs. Synth v1 uses 260 examples, with epoch 50 selected. Recorded warm starts carry Synth v1 into midband RL v2, then through D\&D Character into Phone Number. The release also reports the multi-task Generalized C++ checkpoint. Luna supplies a separately hosted reference evaluation.

Prompt contracts are part of each result. Base and Luna use original prompts; SFT v5 uses fixed26-contract; Synth, Phone Number, and Generalized C++ use fixed26-contract-v2. The contract overlays supply API/build information derived from the tests. These are development-suite evaluations used in the project's improvement process. The Generalized C++ result uses a selected best-four cohort and has six training task IDs in common with Fixed26.

\subsection{Training configuration}

The Phone Number receipt records 20 GRPO updates on eight H100 80 GB GPUs, BF16 execution, tensor parallelism four~\cite{megatron}, and expert parallelism eight. LoRA uses rank 16 and alpha 32. The rollout batch contains eight prompts with 32 responses each; learning rate is \(3\times10^{-5}\), with KL coefficient 0.1. The scored adapter is iteration 14. Appendix~\ref{app:recipe} gathers the full settings and the result-specific evidence checks.

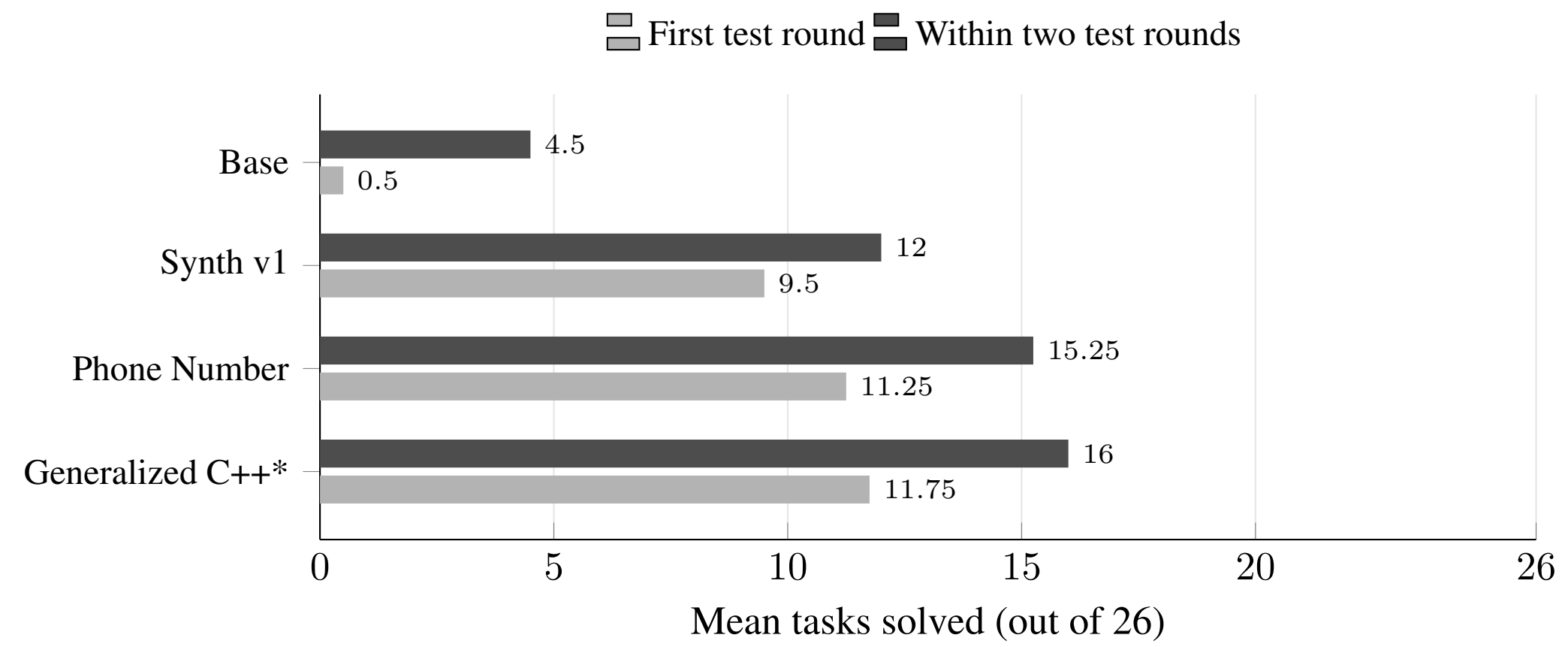
\begin{figure}[htbp]
\centering
\begin{tikzpicture}
\begin{axis}[width=0.88\linewidth,height=5.5cm,xbar,
 xmin=0,xmax=26,xtick={0,5,10,15,20,26},
 xlabel={Mean tasks solved (out of 26)},
 symbolic y coords={Generalized C++*,Phone Number,Synth v1,Base},
 ytick=data,yticklabel style={font=\small},
 enlarge y limits=0.22,bar width=7pt,
 axis x line*=bottom,axis y line*=left,
 xmajorgrids,grid style={black!10},
 legend style={at={(0.5,1.06)},anchor=south,legend columns=2,
 draw=none,font=\small},
 nodes near coords,nodes near coords style={font=\scriptsize},
 every axis plot/.append style={draw=none}]
\addplot[fill=black!30] coordinates {(11.75,Generalized C++*) (11.25,Phone Number) (9.5,Synth v1) (0.5,Base)};
\addplot[fill=black!70] coordinates {(16,Generalized C++*) (15.25,Phone Number) (12,Synth v1) (4.5,Base)};
\legend{First test round,Within two test rounds}
\end{axis}
\end{tikzpicture}

\caption{Mean tasks solved out of 26, from the release results. Four trials per row. *Generalized C++ is the selected best-four cohort. Bars show observed outcomes under the contracts in Section~\ref{sec:contracts}; the complete set of configurations appears below. }
\label{fig:results}
\end{figure}

\begin{table}[htbp]\centering
\caption{Release-reported means and recovery counts. Each row contains four trials and 104 task-trial trajectories. The Generalized C++ row is a development-suite evaluation using test-derived API/build prompt information with the selection and training overlap described in Section~\ref{sec:contracts}. }
\label{tab:t3}
\small\setlength{\tabcolsep}{4pt}\renewcommand{\arraystretch}{1.14}
\begin{tabularx}{\linewidth}{@{}>{\hsize=1.42\hsize\linewidth=\hsize\raggedright\arraybackslash}X>{\hsize=0.86\hsize\linewidth=\hsize\raggedright\arraybackslash}X>{\hsize=0.86\hsize\linewidth=\hsize\raggedright\arraybackslash}X>{\hsize=0.86\hsize\linewidth=\hsize\raggedright\arraybackslash}X@{}}
\toprule

\textbf{Configuration} & \textbf{First round mean / 26} & \textbf{Two rounds mean / 26} & \textbf{Conditional recovery} \\ \midrule

GLM-4.7-Flash base & 0.50 (1.9\%) & 4.50 (17.3\%) & 16/102 (15.7\%) \\

Luna reference & 6.25 (24.0\%) & 16.75 (64.4\%) & 42/79 (53.2\%) \\

SFT v5 & 6.00 (23.1\%) & 10.25 (39.4\%) & 17/80 (21.2\%) \\

Synth v1, epoch 50 & 9.50 (36.5\%) & 12.00 (46.2\%) & 10/66 (15.2\%) \\

Execution-midband RL v1 & 8.25 (31.7\%) & 13.00 (50.0\%) & 19/71 (26.8\%) \\

Execution-midband RL v2 & 10.50 (40.4\%) & 14.50 (55.8\%) & 16/62 (25.8\%) \\

Phone Number, iter 14 & 11.25 (43.3\%) & 15.25 (58.7\%) & 16/59 (27.1\%) \\

Generalized C++, iter 14* & 11.75 (45.2\%) & 16.00 (61.5\%) & 17/57 (29.8\%) \\

\bottomrule
\end{tabularx}
\end{table}

\subsection{Reported completion results}

The first-round mean rises from 0.5 for the base configuration to 6 for SFT v5 and 9.5 for Synth v1. The selected Generalized C++ cohort reaches 11.75 first-round and 16 cumulative solves. These results compare experimental configurations with different prompt contracts.

\subsection{Solving and repairing move differently}

Synth v1 solves more tasks in the first round than SFT v5, yet its conditional recovery is 15.2\%, compared with 21.2\% for SFT v5. The Phone Number checkpoint reaches 27.1\%, and the selected Generalized C++ cohort reaches 29.8\%. Luna's 53.2\% recovery rate provides a useful reference point under its separate hosted setup.

A prepared repair prompt supplies a known problem state and diagnostic context during training. A two-round evaluation asks the agent to react to feedback from its own preceding attempt. We use two-round completion to examine the relationship between practice on prepared repairs and repair after test feedback.

\subsection{Task-level outcomes}

Across four trials, first-round success on phone-number validation changes from 1/4 at Synth v1 to 4/4 at the Phone Number checkpoint. Other tasks move too: Parallel Letter Frequency rises from 0/4 to 3/4, while Gigasecond and Sublist decline. These are checkpoint-level observations across the intervening training lineage.

\begin{table}[htbp]\centering
\caption{Selected task-level outcomes from the released per-task table.}
\label{tab:t4}
\small\setlength{\tabcolsep}{4pt}\renewcommand{\arraystretch}{1.14}
\begin{tabularx}{\linewidth}{@{}>{\hsize=1.05\hsize\linewidth=\hsize\raggedright\arraybackslash}X>{\hsize=1\hsize\linewidth=\hsize\raggedright\arraybackslash}X>{\hsize=1\hsize\linewidth=\hsize\raggedright\arraybackslash}X@{}}
\toprule

\textbf{Task} & \textbf{Synth v1 first-round successes} & \textbf{Phone Number first-round successes} \\ \midrule

Phone Number & 1 / 4 & 4 / 4 \\

Parallel Letter Frequency & 0 / 4 & 3 / 4 \\

Gigasecond & 3 / 4 & 0 / 4 \\

Sublist & 3 / 4 & 1 / 4 \\

\bottomrule
\end{tabularx}
\end{table}

\subsection{Uncertainty estimates}

The released task-bootstrap~\cite{bootstrap} intervals show the variation across this small task suite. Phone Number's first-round mean is 11.25/26 with a 95\% interval of 7.5-15; its cumulative mean is 15.25/26 with an interval of 11-19.25. Generalized C++ has corresponding intervals of 8-15.5 and 11.75-20. These summarize the reported cohorts; selection and prompt-contract differences remain part of the experimental setup.

\section{Discussion}

\subsection{Research decisions and automated operations}

People specify the task, choose the behavior to practice, design checks, and decide which experiment to run. The configured system repeatedly builds candidates, tests their behavior, assigns rewards, updates adapters, and records evaluations.

\begin{table}[htbp]\centering
\caption{Research decisions and repeated operations handled by the configured system.}
\label{tab:t5}
\small\setlength{\tabcolsep}{4pt}\renewcommand{\arraystretch}{1.14}
\begin{tabularx}{\linewidth}{@{}>{\hsize=1\hsize\linewidth=\hsize\raggedright\arraybackslash}X>{\hsize=1\hsize\linewidth=\hsize\raggedright\arraybackslash}X@{}}
\toprule

\textbf{Research and engineering decisions} & \textbf{Repeated operations handled by the system} \\ \midrule

Choose the target behavior and intervention. & Construct the configured task variants and prompts. \\

Define requirements, references, and controls. & Compile, execute, score, and retain check outcomes. \\

Select training settings and checkpoints to study. & Sample responses, update adapters, and run configured evaluations. \\

Interpret results and choose the next experiment. & Produce task-level records for comparison and diagnosis. \\

\bottomrule
\end{tabularx}
\end{table}

\subsection{Future work}

Future work could use recorded failures to support diagnosis and selection of the next training intervention. It could also expand independently implemented task-specific checkers and build curricula that preserve acquired behavior while introducing new mechanisms. Further studies could examine repair using feedback from the model's own attempts and compare interventions on fresh task families under matched evaluation contracts.

\section{Conclusion}

Compiler diagnostics and test outcomes provide feedback for post-training. This project connects that feedback to targeted C++ practice, structured rewards, low-rank model adaptation, and checkpoint-level evaluation. The reported post-training configurations have higher task completion than the base configuration, under differing prompt contracts. The Generalized C++ result uses a selected trial cohort and has training-task overlap with the evaluation suite. Task-level results show improvements and regressions.

\section*{Use of AI Tools}

OpenAI Codex was used to assist with manuscript organization, drafting and revision, literature searches, and LaTeX preparation. The authors reviewed and edited the resulting material and take full responsibility for the final manuscript.

\clearpage

\appendix

\section{Training configuration and reward definitions}
\label{app:recipe}

\subsection{Training configuration}

\begin{table}[htbp]\centering
\caption{Recorded phone-number validation training configuration.}
\label{tab:t6}
\small\setlength{\tabcolsep}{4pt}\renewcommand{\arraystretch}{1.14}
\begin{tabularx}{\linewidth}{@{}>{\hsize=0.72\hsize\linewidth=\hsize\raggedright\arraybackslash}X>{\hsize=1.28\hsize\linewidth=\hsize\raggedright\arraybackslash}X@{}}
\toprule

\textbf{Setting} & \textbf{Recorded value} \\ \midrule

Base model and adaptation & GLM-4.7-Flash; LoRA rank 16, alpha 32. \\

Training hardware & 8 \(\times\) H100 80 GB; BF16; tensor parallel 4; expert parallel 8. \\

Dispatch and serving & DeepEP~\cite{deepep} flex dispatch; SGLang~\cite{sglang} rollout service; Miles~\cite{miles} training integration. \\

Rollout and batch & 8 prompts \(\times\) 32 samples; global batch 256; 20 GRPO updates. \\

Optimization & Learning rate \(3\times10^{-5}\); constant schedule; warmup fraction 0.1; KL coefficient 0.1. \\

Generation & Sequence limit 12,288; response limit 8,192; temperature 0.7; thinking enabled. \\

Warm start and scored checkpoint & Initial adapter: D\&D Character, iteration 19. Evaluated adapter: Phone Number, iteration 14. \\

\bottomrule
\end{tabularx}
\end{table}

These settings come from the released phone-number validation training record and launch configurations. The record also reports wall time and peak memory; timing is marked unverified.

\subsection{Experiment-specific reward definitions}

\begin{table}[htbp]\centering
\caption{Experiment-specific reward definitions and roles.}
\label{tab:t7}
\small\setlength{\tabcolsep}{4pt}\renewcommand{\arraystretch}{1.14}
\begin{tabularx}{\linewidth}{@{}>{\hsize=0.65\hsize\linewidth=\hsize\raggedright\arraybackslash}X>{\hsize=1.35\hsize\linewidth=\hsize\raggedright\arraybackslash}X@{}}
\toprule

\textbf{Reward path} & \textbf{Definition and role} \\ \midrule

Whole-file execution reward & Default scoring: prohibited edits or runtime violations -1; other parse failures -0.8; compilation failure or candidate timeout -0.5; full pass +1; otherwise 0.6 \(\times\) the fraction of passed tests. Recoverable formatting errors incur a penalty of 0.1; scores are clipped to [-1,1]. A strict binary mode is also available. \\

Phone-number validation reward & Mean of 12 signed Boolean check outcomes, in [-1,1]. The recorded outcomes replace the default execution reward. Execution-environment errors are flagged separately. \\

Current shared C++ verification reward & Validated pass +1 (0.975 with a recoverable formatting error). Failed candidates receive -1 plus weighted passed-check credit and available semantic-test credit, clipped to [-1,0]. Invalid execution records are flagged as infrastructure errors. \\

Current task-specific verification adjustment & A positive shared verification score is retained when required task-specific checks pass. Otherwise use \(0.5\min(R_s,0) + 0.5(f_t -1)\). Invalid task-specific evaluations are flagged separately. \\

\bottomrule
\end{tabularx}
\end{table}

In the adjustment formula, \(R_s\) denotes the shared verification score and \(f_t\) the task-specific requirement fraction. The current shared C++ verification reward weights its six scored checks by 0.10, 0.20, 0.20, 0.45, 0.025, and 0.025. These definitions apply to their respective implementations. Historical checkpoints retain the reward configurations recorded for their runs.

\section{Results, artifacts, and reproduction}

\subsection{Trial variability and confidence intervals}

\begin{table}[htbp]\centering
\caption{Trial variability and task-bootstrap confidence intervals for the reported configurations.}
\label{tab:t8}
\small\setlength{\tabcolsep}{4pt}\renewcommand{\arraystretch}{1.14}
\begin{tabularx}{\linewidth}{@{}>{\hsize=0.9\hsize\linewidth=\hsize\raggedright\arraybackslash}X>{\hsize=1.1\hsize\linewidth=\hsize\raggedright\arraybackslash}X>{\hsize=1.1\hsize\linewidth=\hsize\raggedright\arraybackslash}X>{\hsize=0.9\hsize\linewidth=\hsize\raggedright\arraybackslash}X@{}}
\toprule

\textbf{Configuration} & \textbf{First round SD; range; 95\% CI} & \textbf{Two rounds SD; range; 95\% CI} & \textbf{Recovery 95\% CI} \\ \midrule

Base & 0.58; 0-1; 0-1.25 & 1.29; 3-6; 2.25-6.75 & 7.8-24.8\% \\

Luna & 1.89; 5-9; 3.25-9.5 & 1.50; 15-18; 12.75-20.5 & 36.8-70\% \\

SFT v5 & 1.63; 4-8; 3-9.25 & 1.71; 8-12; 7-13.75 & 11.9-31.6\% \\

Synth v1 & 0.58; 9-10; 5.75-13.5 & 0.82; 11-13; 8.25-15.75 & 7.4-24.4\% \\

Midband RL v1 & 2.06; 6-11; 4.75-12 & 0.82; 12-14; 9-17 & 14.5-41.4\% \\

Midband RL v2 & 1.29; 9-12; 7-14 & 2.08; 12-17; 10.5-18.5 & 12.9-41.9\% \\

Phone Number & 0.50; 11-12; 7.5-15 & 0.50; 15-16; 11-19.25 & 13.1-44.2\% \\

Generalized C++* & 1.50; 10-13; 8-15.5 & 1.41; 14-17; 11.75-20 & 14.5-48.9\% \\

\bottomrule
\end{tabularx}
\end{table}

Scores are numbers of tasks solved out of 26; recovery intervals are percentages. Standard deviations (SD) and ranges are across trials. Confidence intervals use 100,000 task-clustered bootstrap resamples, keeping all four trial outcomes together for each sampled task. The Generalized C++ row uses a selected best-four cohort.

\subsection{Reproduction checks}

The run manifests identify the datasets, adapters, evaluation archives, and launch configurations used in each reported experiment.

These commands check the recorded SFT v5 and Phone Number results and artifact identities. Both checks were reported as successful during study preparation. Run them from the release checkout:

\begin{lstlisting}[language=bash]
python results/sft-v5-aiderfmt-1117-4trials/audit/verify.py
python results/phone-number-kernel12-GRPO20/recompute.py --check
\end{lstlisting}

Use each result's recorded launch configuration to reproduce its evaluation. The Generalized C++ run also requires its task registry, admission inputs, and associated run archive.

\subsection{Data and model artifacts}

\begin{itemize}

\item \href{https://huggingface.co/WootzappLab/phone-number-kernel12-GRPO20/tree/adf419fcb32baa335d80c3d9a96c618f3f286a14}{Phone-number validation adapter and run archive}

\item \href{https://huggingface.co/WootzappLab/generalized-cpp-kernel-GRPO20/tree/269560b1f4d6471a1726dd7a95712ef4e3da0838}{Multi-task C++ adapter and selected evaluations}

\item \href{https://huggingface.co/datasets/WootzappLab/glm47-synth-v1-dataset/tree/face23163ca2e9c27f2506b8c757af6ed666dfb7}{Supervised training dataset (Synth v1)}

\item \href{https://huggingface.co/datasets/WootzappLab/glm47-synth-v1-fixed26-evals/tree/ec8b93b8f5916f81b9fecdc9f55960919ef92e2e}{Checkpoint evaluation archive (Synth v1)}

\end{itemize}
\end{document}